# Ductile fracture of HDPE thin films: failure mechanisms and tuning of fracture properties by bonding a rubber layer

Rahul G. Ramachandran[a,1,*], Zachary Kushnir[a], Deepak Langhe[e], Sachin S. Velankar[c,a,d,*], Spandan Maiti[b,a,d]

[a] Department of Mechanical Engineering and Materials Science, University of Pittsburgh, Pittsburgh, PA, USA
[b] Department of Bioengineering, University of Pittsburgh, Pittsburgh, PA, United States
[c] Department of Chemical and Petroleum Engineering, University of Pittsburgh, Pittsburgh, PA, United States
[d] McGowan Institute for Regenerative Medicine, University of Pittsburgh, Pittsburgh, PA, United States,
[e] TD Polymers LLC, Tomball, Texas, United States,

**Abstract**

High-density polyethylene (HDPE) thin films, while inherently ductile, exhibit poor flaw tolerance. Our experiments show that they fail prematurely not at the point of maximum stretch, but at the boundary of a necked region or notch-tip plastic zone. This study investigates this counter-intuitive failure mechanism and demonstrates how an elastomeric interlayer can mitigate it to enhance toughness. Through a combined experimental (uniaxial, center-notched, pure shear tests) and computational approach, we analyze freestanding HDPE and HDPE-SEPS-HDPE trilayer laminates. Finite element simulations reveal that failure in free-standing HDPE is not governed by a maximum stretch criterion. Instead, it is driven by ductile damage, modeled using a damage parameter that depends on stress-triaxiality and plastic strain. This damage localizes initially at the notch or neck center, but migrates into a "hotspot" at the neck/plastic zone boundary, matching the observed crack initiation site. The addition of a soft SEPS interlayer fundamentally alters this behavior by suppressing the ductile damage, causing the failure mechanism to switch from being damage-driven at the neck/plastic zone boundary to being stretch-driven at the notch tip. This switch enhances flaw tolerance and stretchability by relocating and delaying fracture initiation in both defect-free and notched geometries. This work exposes the failure mechanism in HDPE thin films and provides a mechanistically-grounded framework for toughening ductile polymers by manipulating the competition between damage and stretch localization.

## 1 INTRODUCTION

High-density polyethylene (HDPE), one of the most produced man-made materials in the world by volume(Malpass, 2010; Piringer and Baner, 2008), is widely used for disparate applications, including pipe

---

[1] Current address: Institute of Applied Mechanics, Friedrich-Alexander-Universität Erlangen-Nürnberg, Erlangen, Germany
[*] Corresponding authors: rahul.ramachandran@fau.de (R. G. Ramachandran), velankar@pitt.edu (S. S. Velankar)

and fittings (Nguyen et al. 2021), packaging (Piringer and Baner, 2008), and biomedical applications (Fouad and Elleithy, 2011). A key characteristic of polyethylene is its stiff-soft-stiff stress-strain response (see illustration in Figure 1A), which allows it to undergo significant inelastic deformation through cold drawing (Argon, 2013; Carothers and Hills, 1932; Hutchinson and Neale, 1983). During this process, a localized neck forms, stabilizes, and then propagates along the length of the material under tension as illustrated in Figure 1B.

However, while HDPE thin films are inherently ductile, they are also flaw intolerant, failing prematurely in the presence of defects or under high strain rates (see Figure 1C and Section 4.1). The failure of tensile specimen in Figure 1C and notched specimens (shown later) reveal that fracture initiates not in the middle of the neck where the stretch is expected to be maximal, but at the boundary of the highly deformed necked region. To our knowledge, this specific failure mode at the neck boundary has not been previously reported in the literature. Consequently, failure criteria based solely on maximum stress, stretch (Kattekola et al., 2013), or plastic strain (Gearing and Anand, 2004), commonly used for ductile materials, are insufficient to predict this behavior, highlighting a significant gap in the mechanistic understanding of HDPE thin film fracture.

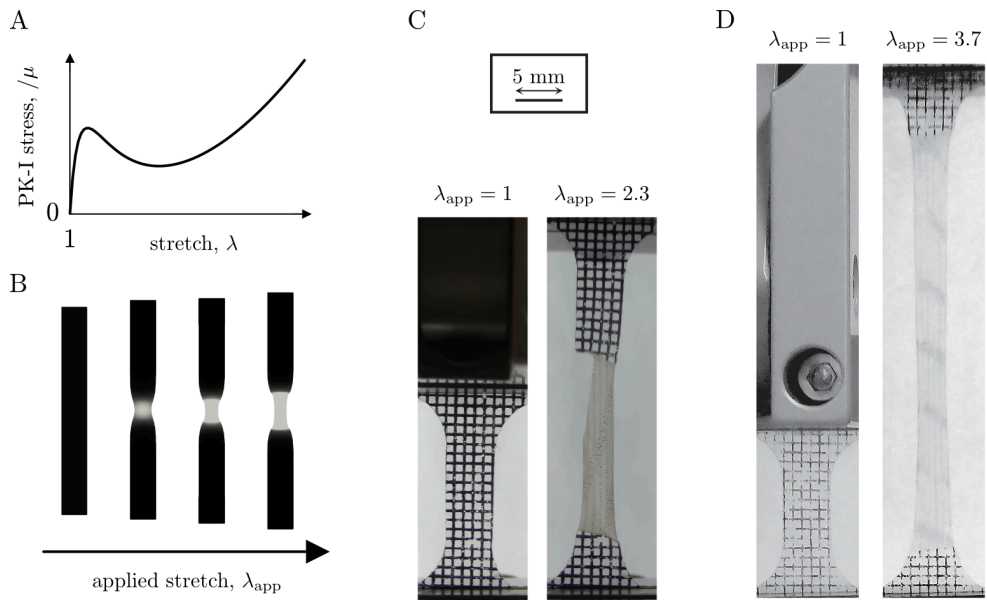

Figure 1: Tensile behavior and failure of HDPE thin films. (A) A representative non-monotonic stress-stretch curve of a stiff-soft-stiff material. The first Piola–Kirchhoff (PK-I) stress is normalized by the shear modulus ($\mu$). (B) Schematic illustration of cold drawing (Hutchinson and Neale, 1983; Ramachandran et al., 2020) of a material with constitutive behavior shown in Figure 1 A. As the applied stretch ($\lambda_{\text{app}}$) increases, a neck forms and propagates. (C) Experimental result for a standalone HDPE film stretch at a displacement rate of 1200 mm/min, showing premature fracture at the boundary of the necked region. (D) HDPE-SEPS-HDPE trilayer with rubber fraction $\zeta = 0.15$ undergoes cold drawing without failure, demonstrating enhanced flaw tolerance.

In this work, we further demonstrate that bonding a soft elastomeric interlayer, specifically Styrene-Ethylene-Propylene-Styrene (SEPS), to the HDPE film suppresses this premature failure, allowing the composite to stretch without fracture and introducing flaw tolerance (Figure 1D). Accordingly, the objectives of this paper are twofold: first, we investigate mesoscopic failure mechanisms of HDPE thin films, and subsequently, we study how bonding a rubber layer modifies these mechanisms. Ultimately, our goal is to develop HDPE-rubber composite films with tunable failure properties in a mechanistically informed manner.

To achieve these objectives, we employ a combined experimental and computational approach. We conducted experiments on free-standing HDPE thin films and HDPE-SEPS-HDPE trilayers of varying SEPS thickness fractions to capture their failure behavior. Tensile experiments were conducted on center-hole-notched (CHNT) and edge-notched tensile (ENT) specimens, concurrently tracking failure. The size of the notches was varied to study flaw tolerance. Additionally, we conducted fracture tests on pure-shear (PS) fracture specimens while concurrently recording the notch tip deformation to capture crack initiation. Notched geometries allow experiments at slow speeds as well as localizes the region of interest to a small region around the notch tip; allowing for easy visual analysis. Meanwhile, different notch geometries quantify the effect of stress concentration on the failure behavior.

Next, we performed finite deformation elasto-plastic finite element analysis to quantify the stress state in the "defect-free" tensile specimens as well as notched, CHNT and PS fracture specimens across free-standing HDPE and HDPE-SEPS-HDPE trilayer composites with varying SEPS thickness fractions. By correlating our experimental observations of crack initiation with computational predictions based on a stress-triaxiality and plastic strain dependent damage criterion, we identified the key mechanical fields driving failure.

Ultimately, this work provides a mechanistically-informed framework for designing tougher and more reliable HDPE-based composites. We demonstrate how the failure properties of these films can be tuned by varying the properties and thickness of the bonded rubber layer. This paper is organized as follows: Section 3 details the experimental and computational methods. Section 4 presents the experimental and simulation results, contrasting the behavior of standalone and composite films. In Section 5, we discuss the failure mechanisms and the role of the rubber layer in tuning fracture properties. We conclude in Section 6 with a summary of our findings and their broader implications.

## 2 List of symbols and definitions

| | |
|---|---|
| CHNT specimen | Center-hole-notched tensile specimen |
| ENT specimen | End-notched tensile specimen |
| PS fracture specimen | Pure-shear fracture specimen |
| $\lambda, \lambda_{\text{app}}, \lambda_{\text{neck}}$ | Local uniaxial stretch, applied stretch, neck stretch |
| PK-I stress | First Piola-Kirchhoff stress |
| Nominal stress | Reaction force / Cross section area of the gauge section, $F/A$ |
| $\mu$ | Shear modulus |
| $\zeta$ | Rubber thickness fraction in the composite |
| $\delta$ | Cross head displacement |
| $L$ | Length of the gauge section |
| $\boldsymbol{F}, \boldsymbol{C} = \boldsymbol{F}^T \boldsymbol{F}$ | Deformation gradient, Right Green Cauchy strain tensor |
| $I_1, I_2$ | First and second invariants of Right Green Cauchy strain tensor |
| $J = \det(\boldsymbol{F})$ | Jacobian |
| $\kappa$ | Internal penalty parameter to penalize volume changes |
| $C_1^r, C_2^r, C_3^r$ | Material parameters for the SEPS |
| $C_1^a, C_2^a, C_1^c$ | Material parameters for the amorphous and the crystalline phase for the HDPE |
| $\psi^r, \psi^a, \psi^c$ | Strain energy density function for the rubber, amorphous phase and crystalline phase for HDPE |
| $\boldsymbol{F}_e \boldsymbol{F}_p$ | Elastic and Plastic part of the deformation gradient |

| $S_e, C_e, M_e, M_e^d$ | Elastic second Piola–Kirchhoff stress, elastic Right Green Cauchy strain tensor, elastic Mandel stress, and the deviatoric part of the elastic Mandel stress |
|---|---|
| $\epsilon_p, \sigma_y, H$ | internal plastic parameter, yield stress and the linear strain hardening parameter for the HDPE |
| $c, d$ | Length of the sharp edge notch and diameter of the center hole notch |
| $w$ | Width of the gauge section of the tensile specimen |
| $\ell$ | Crack ligament length |
| $D$ | Ductile damage parameter |
| $\max(D)$ | Maximum of ductile damage parameter |
| $\lambda_1$ | Primary principal stretch |
| $\max(\lambda_1)$ | Maximum of primary principal stretch |
| $\chi$ | Axial position of $\max(D)$ in tensile specimen from the center. |
| $\theta$ | Azimuthal position of $\max(D)$ from the notch tip in CHNT and PS fracture specimen |

## 3 METHODS

### 3.1 Experimental Methods

HDPE and HDPE-SEPS-HDPE trilayer sheets with various rubber thickness fractions $\zeta$ were prepared by continuous extrusion in a multilayer coextrusion die by Polymer Plus, Akron, OH. Briefly, three extruders were used to feed a three-layer feed block and the resulting films were collected on a chill roll. Various thickness ratios were targeted by controlling the flow rate of the extruders. However, the rheological properties of the HDPE and the SEPS selected were not well-matched for coextrusion (specifically the SEPS has a much higher viscosity than HDPE). Accordingly, there was some variation in the relative thickness of the layers across the width of the films, and across the length of the rolls. Accordingly, the rubber content of each testing specimen was measured separately. This was done by taking a sample from the film immediately adjacent to the specimen, measuring its total thickness, dissolving the

SEPS in toluene, and then measuring the thickness of the two face-layers of plastic. The ratio of the rubber thickness to the total thicknesses was defined as rubber thickness fraction, $\zeta$.

To characterize the properties of SEPS, free-standing SEPS sheets were solvent cast out of toluene solution using a micrometer-adjustable film spreader (Gardco). In either case, dog bone-shaped tensile samples with gauge length 20 mm and width 6 mm were then cut with die, with the gauge section aligned in the machine-direction. Tensile experiments were conducted on each of the two pure materials as well as the trilayer composites. Tensile tests were performed on undamaged dog bone-shaped specimens as well as sharp-edge notched, or center-hole-notched specimens. Sharp notches of different length were cut into the edge of dogbone samples using razor blades while monitoring depth with an optical microscope, whereas the holes were cut with round punches of different diameters. All tensile tests were done on an MTS universal tensile testing platform. The samples were mounted and stretched with a grip velocity of 20 mm/min. Pure-shear fracture tests were performed on 25 mm wide and 10 mm long samples (grip to grip) at 5mm/min. The initial sharp notches were made with a razor blade and had a length of 10 mm.

For all the tests, the reaction forces ($F$) and cross head displacement ($\delta$) data were collected from the tensile testing machine. For the tensile specimens, reaction force is normalized with the original cross section area ($A$) of the gauge section to get the nominal stress, $F/A$. The stretch was calculated from cross head displacement ($\delta$) and length of the gauge section ($L$) as $\lambda_{\text{app}} = 1 + \delta/L$. For selected tests, a camcorder recorded the deformation of the whole specimen, while a Dino-lite® digital microscope was used for close up visual recording of the notch tip deformations. Black glitter particles and sharpie ink were applied on the surface of the specimen for better visualization of the large deformations. We also performed digital image correlation for true stretch quantification of the deformation in the neck of the tensile specimen as well as at the notch tip using custom Digital Image Correlation (DIC) code. The details are described previously (Ramachandran et al., 2018).

### 3.2 Material model

The SEPS rubber is modelled using a three-parameter - incompressible, rate independent hyperelastic model. The strain energy density function is described as a function of the first ($I_1 = tr(\boldsymbol{C})$) and second ($I_2 = \frac{1}{2}[(tr(C))^2 - tr(\boldsymbol{C})^2]$) invariant of the Right Green Cauchy strain tensor ($\boldsymbol{C} = \boldsymbol{F}^T \boldsymbol{F}$) as in Eq. 1.

$$\psi^r = C_1^r(I_1 - 3) + C_2^r(I_2 - 3) + C_3^r(I_1 - 3)^2 + \frac{\kappa}{2}(J - 1)^2 \qquad (1)$$

Here, $J = \det(\mathbf{F})$ is the determinant of the deformation gradient, and $\kappa$ is an internal penalty parameter to penalize volume changes. A suitably large value for $\kappa$ was taken to model the material as nearly incompressible. The constitutive parameters, $C_1^r$, $C_2^r$ and $C_3^r$ were evaluated by fitting the simulated nominal stress response to the experimental response. These parameters were obtained by fitting the prediction of Eq. 1 for uniaxial tension to experimental data, we estimated $C_1^r = 0$, $C_2^r = 1$ MPa and $C_3^r = 8$ kPa (see fit in Figure 2).

The HDPE is modelled assuming that the mechanical response is a sum of the contributions from the amorphous and crystalline phases (van Dommelen et al., 2003). The amorphous phase was modelled using an incompressible rate independent Gent hyperelastic strain energy density function given in Eq. 2.

$$\psi^a = C_2^a \left[ J_m \ln\left(1 - \frac{I_1 - 3}{J_m}\right) \right] + \frac{\kappa}{2}(J - 1)^2 \tag{2}$$

The crystalline phase was modelled as a rate independent elastic-plastic material. The elastic portion was modelled using an incompressible NeoHookean strain energy density function:

$$\psi^c = C_1^c(I_1 - 3) + \frac{\kappa}{2}(J - 1)^2 \tag{3}$$

The kinematics of plastic deformation was implemented by multiplicative decomposition of the deformation gradient into elastic and plastic components, $\mathbf{F} = \mathbf{F}_e \mathbf{F}_p$. The subscript $e$ and $p$ represent the elastic and plastic parts, respectively. The plastic flow was defined by the isotropic linear hardening yield function, Eq. 4.

$$f(\mathbf{M}_e, \epsilon_p) = \sqrt{\frac{3}{2} \mathbf{M}_e^d : \mathbf{M}_e^d} - [\sigma_y + H\epsilon_p] \leq 0 \tag{4}$$

Here, $\mathbf{M}_e^d$ is the deviatoric component of the Mandel Stress defined as $\mathbf{M}_e = \mathbf{C}_e \mathbf{S}_e$ where, $\mathbf{C}_e = \mathbf{F}_e^T \mathbf{F}_e$, $\mathbf{S}_e = \frac{\partial \psi^c(\mathbf{C}_e)}{\partial \mathbf{C}_e}$ are the elastic part of right Green Cauchy strain tensor and second Piola-Kirchhoff Stress tensor respectively. And $\epsilon_p$, $\sigma_y$ and $H$ are the internal plastic parameter, yield stress and the linear strain hardening parameter.

The constitutive parameters for HDPE were identified by fitting the simulated nominal stress response with the experimental response (see fit in Figure 2). These values are $C_1^c = 35$ MPa, $\sigma_y = 21$ MPa, $H = 20$ MPa, $C_2^a = 100$ kPa, and $J_m = 7.5$.

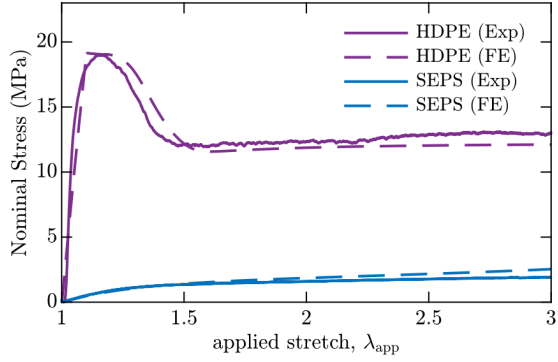

Figure 2: Finite Element (FE) simulated and experimentally (exp) measured nominal stress-stretch plots for HDPE plastic and SEPS rubber where solid lines are experimental measurements from tensile tests on dog bone-shaped samples and dashed lines are simulation results.

### 3.3 Simulation methodology and data postprocessing

Finite element simulations were conducted to study the stress state in tensile loading of rectangular geometries with or without notch (center hole or edge notch). A custom highly parallelized 3D quasistatic nonlinear finite element code was used. It was used previously to study neck propagation (Ramachandran et al., 2020) and wrinkling (Yang et al., 2017) in rubber-plastic laminates. The finite element meshes are shown in Figure 3A and B. Exploiting the symmetry of the geometries, only the one-eighth (symmetric about X-X,Y-Y, and Z-Z) of the rectangular tensile specimen (with and without center hole notch, Figure 3 A, B) and one-quarter (symmetric about Y-Y, and Z-Z) of the rectangular specimen with an edge notch were modeled (Figure 3 C). Roller boundary conditions were imposed on the symmetry planes.

Single-layer geometries of 40 $\mu$m thickness of the three cases were created to simulate free-standing SEPS and HDPE behavior. Trilayer geometries with different SEPS fractions were created for all three cases while keeping the total thickness constant at 40 $\mu$m. For geometries with defects, the edge notch tip radius was 25 $\mu$m and the radius of the center hole was set to 200 $\mu$m. All the meshes had at least two elements per layer along the thickness direction (not shown in Figure 3).

The tensile geometry was meshed using 26820 3D hex elements. For the tensile geometry, the nodes along the symmetry plane along the length were displaced by 0.5 $\mu$m in the thickness direction (-Z), to create a defect and hence ensure consistent localization at the symmetry plane along the length of the geometry. The mesh was refined at the middle of the specimen in the Y-direction to capture the necking and the sharp boundary of the necked region (Figure 3A).

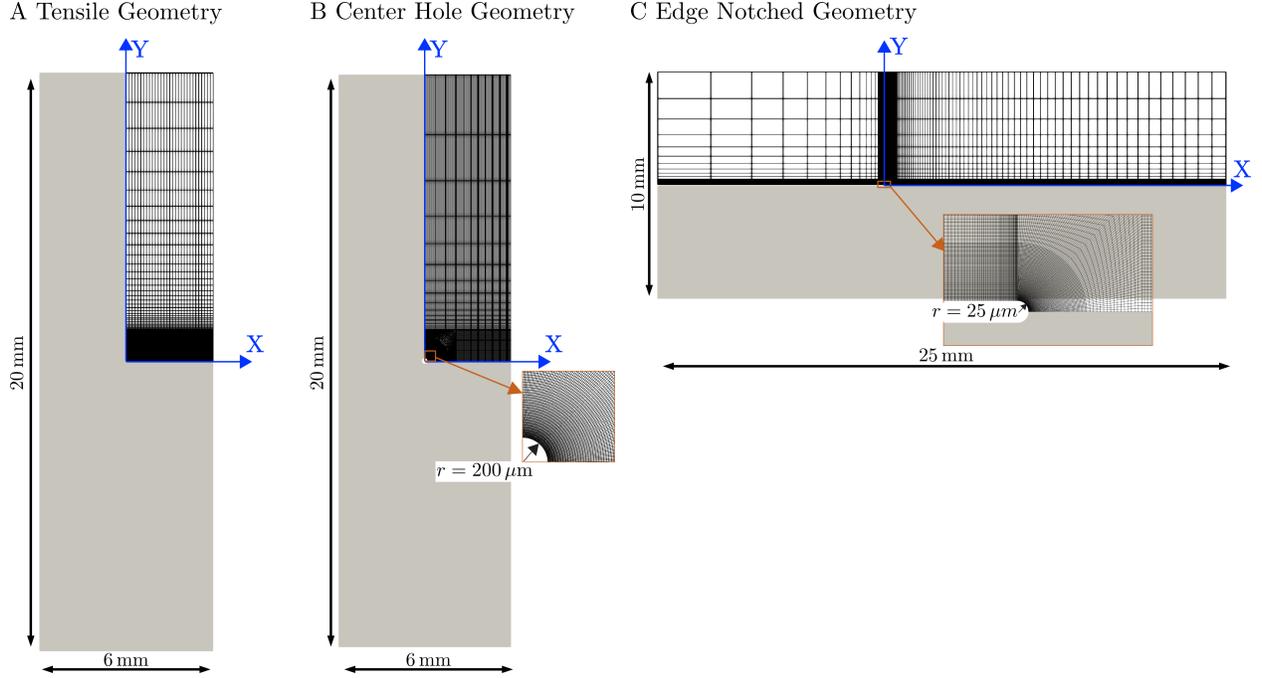

Figure 3: Rectangular geometries of simulations (a) without defects, (b) with a notched center hole (C) with an edge notch. The stretching direction is in the Y-direction.

For the other geometries, the center hole geometry was meshed using 44130 3D hex elements. The mesh was refined around the hole (inset of Figure 3B) as well as ahead of the hole in the X-direction. The edge notched geometry was meshed using 49396 3D hex elements. The mesh was refined at the notch tip and ahead of the notch in X-direction to capture the plastic localized deformation seen ahead of defects (inset of Figure 3C). From the finite element solution, the stretch at each node were extrapolated and recorded.

For quantifying the damage in the HDPE layer, here we will use a ductile damage parameter originally derived to quantify the rate of void growth in metals. Foundational works on ductile fracture has examined the rate of growth of spherical voids (Rice and Tracey, 1969) and cylindrical holes (McClintock, 1968) in plastic materials and was found to be proportional to exponential of stress triaxiality factor $\eta$,

$$\eta = \sigma_m/\sigma_{von} \tag{5}$$

and plastic strain $\epsilon_p$, where $\sigma_m$ is the mean stress and $\sigma_{von}$ is the effective Von Mises stress. The damage parameter $D$ is defined as,

$$D = \exp\left(\frac{3}{2}\eta\right)\epsilon_p \tag{6}$$

Furthermore, principal stretches are evaluated by taking the square root of the eigen values of the Right Green Cauchy tensor $\boldsymbol{C} = \boldsymbol{F}^T\boldsymbol{F}$. The maximum of the principal stretches is called the primary principal stretch $\lambda_1$:

$$\lambda_1 = \max(\text{eigen}(\sqrt{\boldsymbol{C}})) \tag{7}$$

The $\lambda_1$ and $D$ are evaluated at each of the 8 Gauss points in the brick element of the Finite element mesh and interpolated to the nodes and recorded.

## 4 RESULTS

### 4.1 Tensile behavior of "defect-free" HDPE thin film specimens and HDPE-SEPS-HDPE trilayer thin film specimens

We first examine the tensile behavior of nominally "defect-free" specimens (i.e., without deliberately introduced notches or holes) at a slow displacement rate of 20 mm/min. Figure 4A plots the engineering stress versus applied stretch ($\lambda_{app}$) from tensile experiments on free-standing HDPE, SEPS, and HDPE-SEPS-HDPE trilayers with various rubber fractions ($\zeta$). In freestanding HDPE specimens ($\zeta = 0$), yielding and subsequent necking initiate at approximately 15% strain, coinciding with the peak in the engineering stress (Figure 4A), followed by neck propagation that corresponds to a plateau in nominal stress. For the trilayers, the peak stress and the subsequent drop to the stress plateau become shallower with increasing $\zeta$. In contrast, the SEPS has no peak or localization behavior, as its engineering stress rises monotonically with $\lambda_{app}$.

The stretch in the neck of the specimen is measured by digital correlation analysis of the videos as described previously (Ramachandran et al., 2018), is plotted in Figure 4B. The stretch in the neck (defined as maximum local stretch within the specimen), measured at an applied stretch of 2.5, decreases with increasing $\zeta$. Qualitatively, the behavior shown in Figure 4 is similar to that of LLDPE plastics and LLDPE-SEPS bilayer discussed previously (Ramachandran et al., 2020, 2018).

Next we analyze failure in a "defect-free" HDPE tensile specimen; failure is induced by stretching at a displacement rate of 1200 mm/min, exploiting strain rate sensitivity of HDPE (Dasari and Misra, 2003; Manaia et al., 2019). The tensile specimen undergoing failure was shown in Figure 1C. The HDPE film at first showed cold drawing behavior and deformed in tension by necking and subsequent neck propagation. However, we found that the neck could not propagate throughout the length of the gauge section of the dogbone specimen. Instead, after an initial neck propagation up to an applied stretch $\lambda_{app} = 2.1$, the HDPE thin film fractured close to the boundary separating the necked region from the rest of the specimen. This partial neck propagation is reflected as a plateau and the fracture by the rapid decrease in nominal stress in the nominal stress - applied stretch plot for HDPE in Figure 4C.

In contrast HDPE-SEPS-HDPE trilayer of 40 μm thickness ($\zeta = 0.15$) stretched without failure at the 1200 mm/min (Figure 1D). The composite film was able to form a neck that propagated throughout the gauge section of the specimen without failure. This enhanced stretchability is reflected in its stress-stretch curve (Figure 4C, dashed line), which exhibits a prolonged stress plateau extending to a much larger applied stretch.

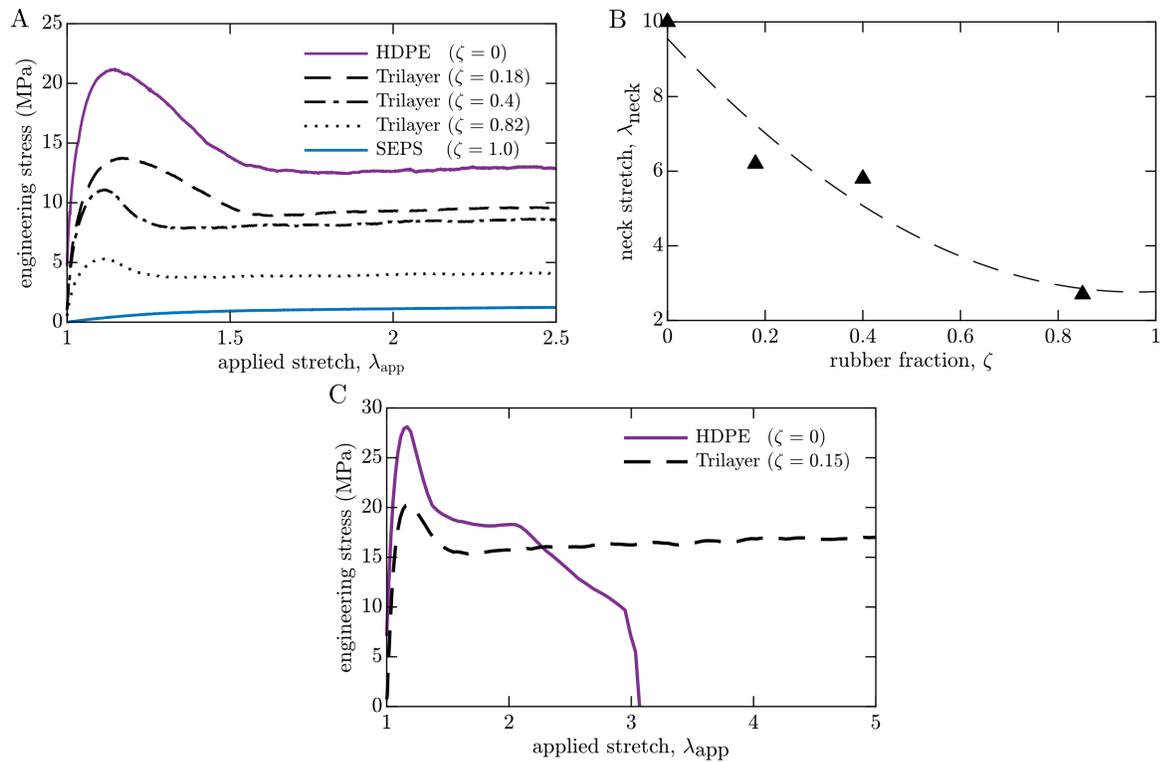

Figure 4: (A) Engineering stress – stretch ($\lambda$) behavior of HDPE and HDPE-SEPS trilayer of various rubber fraction ($\zeta$). (B) The neck-stretch of HDPE and HDPE-SEPS trilayers of various rubber fraction ($\zeta$) measured at an applied stretch of 3. The dashed curve plotted is only a guide to the eye and has no physical significance. (C) The engineering stress vs applied stretch ($\lambda_{app}$) for the HDPE and HDPE-SEPS trilayer ($\zeta = 0.15$) experiments in Figure 1C and D.

## 4.2 Tensile behavior of notched HDPE thin film specimens and HDPE-SEPS-HDPE trilayer thin film specimens

To test the stretchability of HDPE, SEPS, and HDPE-SEPS-HDPE trilayer films with various rubber fractions ($\zeta$), we prepared tensile specimens with controlled defects and stretched them to an applied stretch of $\lambda_{app} = 3$. Figure 6 presents flaw tolerance maps for specimens with edge notches (length $c$) and center holes (diameter $d$). These maps indicate whether a sample stretched without failure (green circles) or failed (red 'x' symbols) for a given defect size and rubber fraction.

Freestanding HDPE ($\zeta = 0$) showed no flaw tolerance for the defect sizes tested, failing with edge notches exceeding 0.1 mm or center holes exceeding 0.12 mm. In contrast, freestanding SEPS ($\zeta = 1$) was highly flaw tolerant, failing only with large edge notches ($> 0.4$ mm) and resisting failure from all tested hole sizes. The trilayer composites exhibited intermediate behavior, in which even the smallest rubber fraction ($\zeta \approx 0.2$) significantly improved flaw tolerance. The defect size required to cause failure increased monotonically with $\zeta$ for both defect types. Overall, bonding a rubber layer made the HDPE flaw-sensitive - its failure now dependent on defect size - while systematically improving its tolerance with increasing rubber thickness. The mechanics at the defect tip are examined in the following sections.

Visual observation of failure revealed that necking strongly influenced crack initiation, especially at low $\zeta$. A highly stretched region, which we denote as the "plastic zone," formed at the notch tip and then either spread from its boundaries or expanded forward. To visualize deformation near the defect, we performed a two-step experiment. The results shown here are for specimens with center holes; similar behavior for edge notches is documented in supplementary Figure S1.

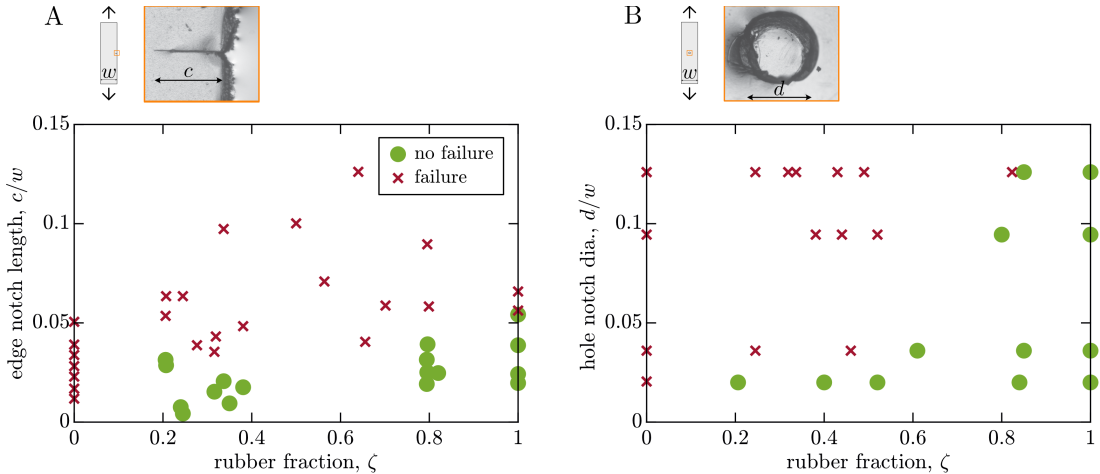

Figure 5: Flaw tolerance map showing whether a sample failed for a given defect size and rubber fraction $\zeta$. (A) Edge notch of length $c$ normalized by gauge width $w$. (B) Center hole of diameter $d$ normalized by $w$.

In the first step, ink was applied around the hole in the undeformed specimen, and the sample was stretched to form a plastic zone, highlighted by the lightened ink color (Figure 6A1–D1). In the second step, the test was paused (Figure 6A2–D2), new ink was reapplied to the plastic zone, and then stretching was resumed (Figure 6A3–D3). This method distinguishes whether additional deformation is accommodated by expanding the plastic zone or by further stretching the material within it.

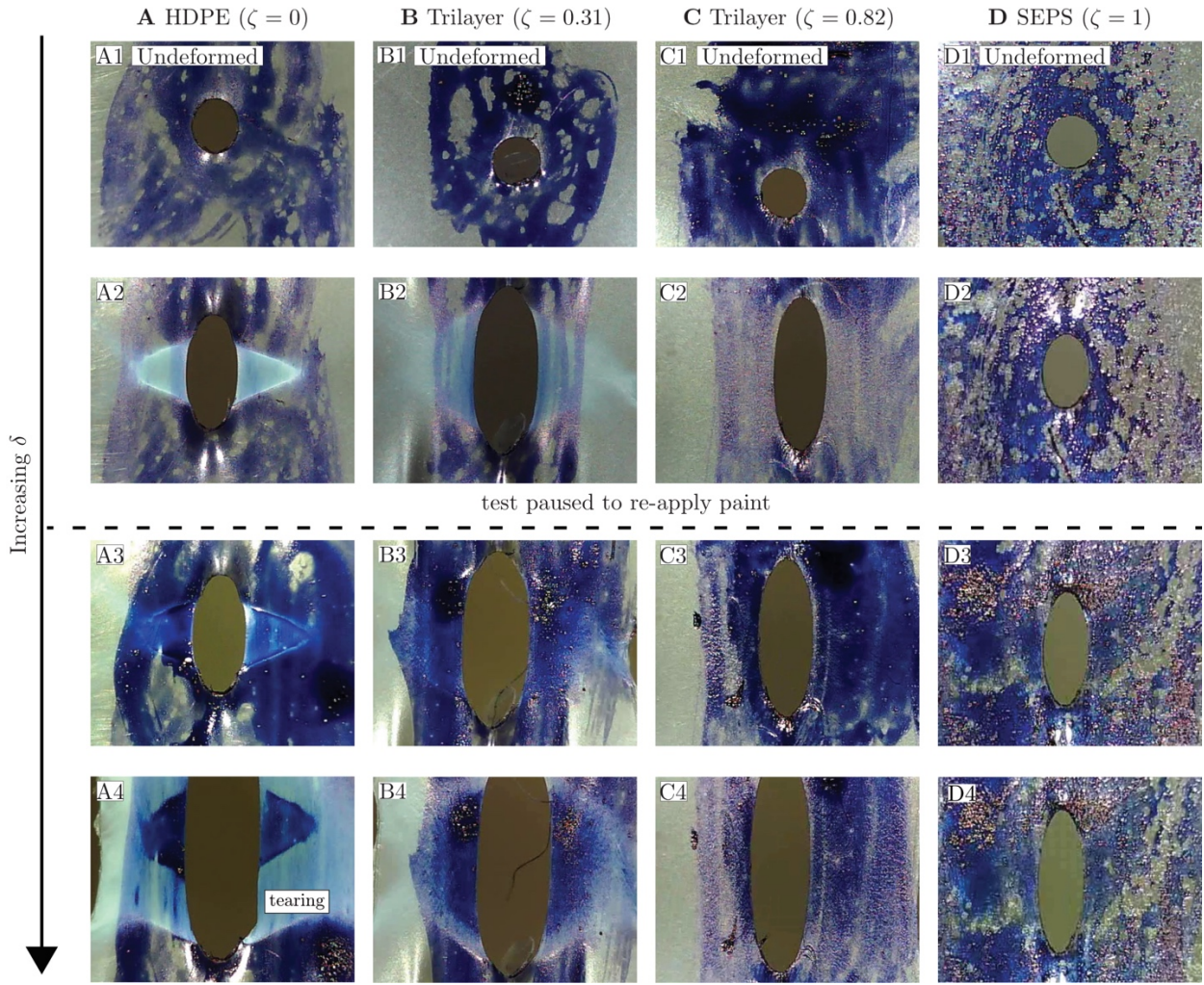

Figure 6: Plastic zone evolution in HDPE-SEPS trilayer films with SEPS thickness fractions, $\zeta = 0$ (free-standing HDPE), 0.31, 0.82, and 1 (free-standing SEPS), captured in a two-step experiment on a tensile specimen with a center hole. (A1–D1) Undeformed configuration. (A2–D2) A small crosshead displacement was applied to form a plastic zone. (A3–D3) The test was then paused, and paint was reapplied to the material within the plastic zone. (A4–D4) Evolution of the plastic zone upon further stretching. The images in each column are sequential; however, the images in a given row were not taken at the same crosshead displacement and should not be directly compared.

For the HDPE film, triangular plastic zones developed and propagated outwards from the hole (Figure 6A2). After re-inking, this zone spread by drawing in new material - similar to neck propagation (Figure 1C) - while the originally stretched material remained largely undisturbed, as shown by the sharp interface between the new and smeared ink (Figure 6A4). A crack eventually initiated at the plastic zone boundary and propagated. Conversely, the free-standing SEPS specimen showed no distinct plastic zone or crack initiation up to the maximum stretch examined (Figure 6D1-4).

The composite with $\zeta = 0.31$ behaved similarly to HDPE, but its triangular plastic zone lacked sharp boundaries (Figure 6B). After re-inking, deformation was accommodated mainly by drawing in new material, though slight smearing of the new ink indicated some additional stretching within the original plastic zone. At $\zeta = 0.82$, the behavior closely resembled that of SEPS: no clear plastic zone formed, deformation occurred broadly around the hole, and no cracks initiated within the stretch examined (Figure 6C).

Similar two-step visualizations for specimens with edge notches yielded substantially similar results (Figure S1). These qualitative experiments motivated the more quantitative study of plastic zone deformation in Sections 4.3 and 4.4.

### 4.3 Pure shear fracture tests of HDPE, SEPS and HDPE-SEPS-SEPS trilayer thin films

This section examines the fracture behavior of a free-standing HDPE film and how it is modified by bonding a rubber layer. We conducted mode I pure shear fracture tests to quantify the forces during the failure of SEPS, HDPE, and HDPE-SEPS-HDPE trilayers. Deformation of the whole specimen was captured via low-magnification video. The experiments used rectangular specimens (25 mm wide) with sharp, 10 mm edge notches and an initial clamp separation of 10 mm. A straight ink line on the crack plane and additional surface markers were used to visualize deformation within and outside the plastic zone, respectively. A typical undeformed specimen is shown in Figure 7A.

Figure 8B shows the normalized force-displacement fracture curves for the HDPE, SEPS, and trilayer films. The force is normalized by the initial ligament area ($A_l$), defined as the product of the specimen's thickness and ligament length ($l$ in Figure 7A). In all cases, the normalized force initially increased before decreasing with deformation. Figure 7C–F show images of the specimens at applied crosshead displacements ($\delta$) of 2.5, 7.5, 10, 12.5, and 15 mm.

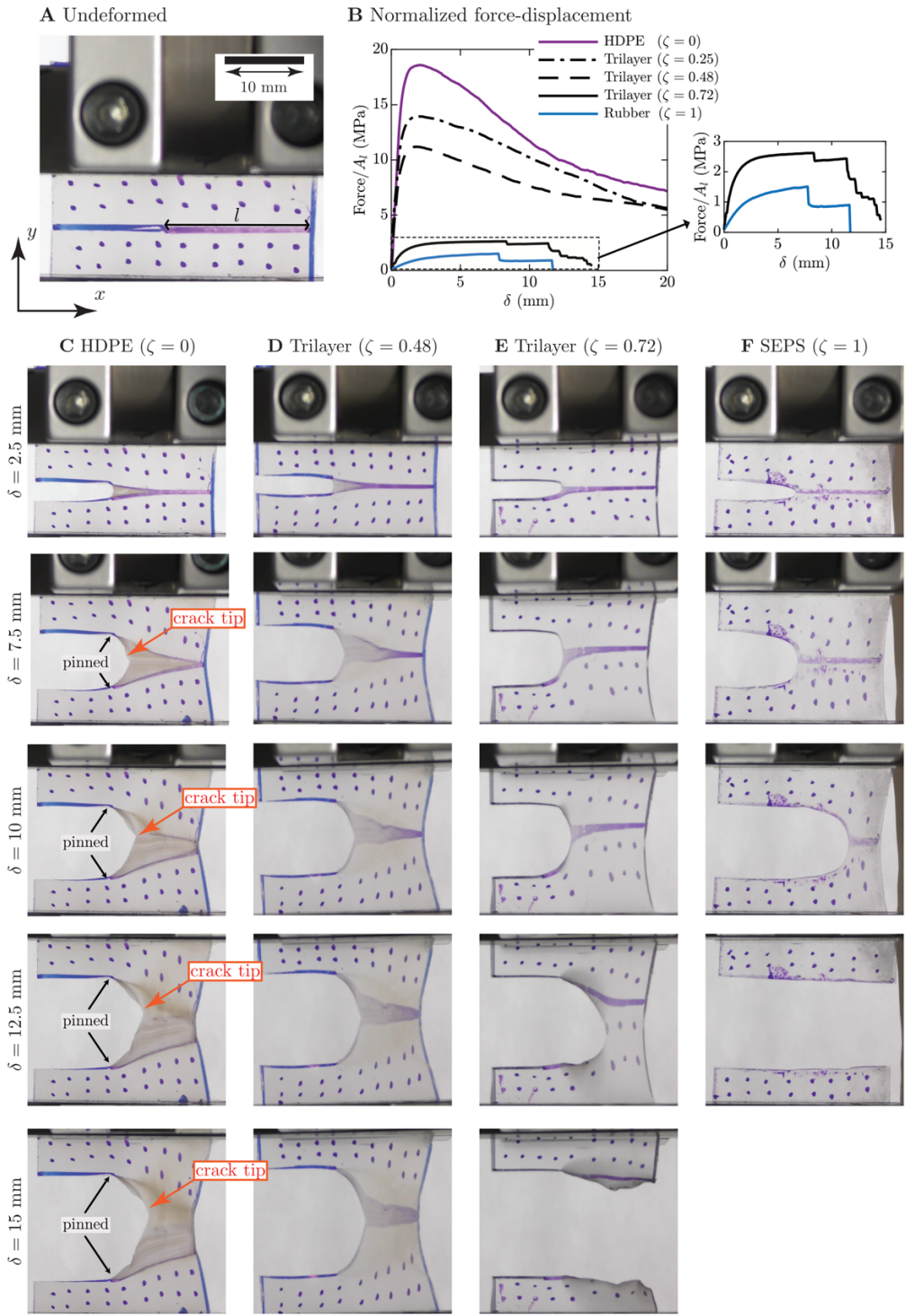

Figure 7: Pure shear fracture experiments on HDPE, SEPS, and HDPE-SEPS-HDPE trilayers with $\zeta$ values of 0.48 and 0.72. (A) The undeformed configuration of a fracture specimen. Only the HDPE specimen is shown, as all other samples appeared similar initially. (B) Normalized force-displacement curves for the fracture of HDPE, SEPS, and HDPE-SEPS trilayer films with $\zeta$ values of 0.25, 0.48, and 0.72. For clarity, the curves for the trilayer with $\zeta = 0.72$ and for SEPS are enlarged and shown separately. (C–F) Images of the fracture process for HDPE, SEPS, and trilayers with $\zeta$ values of 0.48 and 0.72 at various applied displacements, $\delta$. Each row corresponds to a different applied crosshead displacement with a value of $\delta$ = 2.5, 7.5, 10, 12.5, or 15 mm.

The free-standing HDPE film (Figure 7C) fractured as follows. At δ = 2.5 mm (near the maximum force for HDPE), a triangular plastic zone formed at the notch tip. Within this region, severe stretching is indicated by a lighter hue. The transition from the plastic zone to the external region is sharp, similar to the necked region in tensile deformation (Figure 1C). Increasing the displacement to 7.5 mm caused the plastic zone to grow, predominantly in the x-direction. Failure occurred by crack propagation in the plastic zone, accompanied by a gradual decrease in force.

In contrast, the SEPS rubber (Figure 7F) showed different behavior. For $\delta$ up to 7.5 mm, the notch opened progressively without crack initiation or an apparent plastic zone. At δ = 7.7 mm, a crack initiated and propagated unstably until it was arrested near the specimen's edge. The normalized force increased with displacement until the crack initiated, causing a rapid force drop (Figure 8B, enlarged plot). Further displacement led to a complete, unstable fracture at $\delta$ = 11.6 mm.

Turning to the composites, Figure 7B shows their normalized force curves are of a lower magnitude than that of HDPE. At a low rubber-to-plastic ratio ($\zeta$ = 0.25), the fracture behavior was qualitatively similar to HDPE, involving the formation of a plastic zone and crack propagation within it. At $\zeta$ = 0.48 (Figure 7D), the behavior also remained similar to HDPE; however, the plastic zone edges were more diffuse, and the material far ahead of the crack was deformed to a greater extent.

Conversely, the composite with a large rubber fraction ($\zeta$ = 0.72, Figure 7E) showed behavior visually similar to SEPS, but the deformation ahead of the notch tip was more diffuse, as evidenced by the widening of the marker line at $\delta$ = 7.5 mm. Similar to SEPS, the failure was unstable; however, the crack propagated incrementally in small bursts, which appear as small drops and plateaus in the force curve (Figure 8B, magnified view).

Several noteworthy features of the HDPE fracture are apparent. First, the gradual loss of load-carrying capacity with increasing deformation indicates stable crack propagation, although trends in crack propagation were not investigated further in this study. Second, deformation was almost completely localized within the plastic zone, evidenced by the lack of relative displacement of the ink markers outside this zone (Figure 8C). Additionally, material within the plastic zone became pinned at its boundary during crack propagation (marked in Figure 8C). In the composite, increasing $\zeta$ led to more diffuse deformation ahead of the notch tip, which was distributed over a larger volume. Moreover, as $\zeta$ increased, the transition between the highly stretched plastic zone and the bulk material became smoother. Finally, for a large value of $\zeta$, the fracture behavior became rubber-like, as shown by the load curve in Figure 8B.

## 4.4 Notch tip deformation in fracture tests: notch blunting and delaying crack initiation in HDPE SEPS and HDPE-SEPS-HDPE trilayer thin films

Crack initiation in the defect-free HDPE tensile specimen (Figure 1C) and the center-hole-notched specimen (Figure 6A) occurred at the boundary of the necked region and the neck-like notch-tip plastic zone, respectively. In contrast, this behavior was not observed in the HDPE fracture specimen shown in Figure 7C, where crack appeared to initiate in the middle of the neck-like plastic zone.

Therefore, we performed a more detailed visualization of the fracture process to address two points. The first was to quantify the magnitude of deformation around the notch. The second was to differentiate between notch blunting (the opening of the notch without crack initiation) and crack initiation. The experiments from the previous section were repeated, but with video recorded at a high magnification focused on the notch tip. Furthermore, solid marker particles were dispersed on the specimen surface to compute stretch from the relative displacement of the markers, as done previously (Ramachandran et al., 2018). Due to the higher magnification, only the early stages of stretching, up to $\delta = 4.5$ mm, could be followed. The frames in Figure 8 show either notch blunting or crack propagation. All frames in which cracks are propagating were identified from the corresponding videos and are marked with an orange background.

The deformation at the HDPE notch tip (Figure 8A) shows that the formation of a neck-like plastic zone first blunted the notch up to $\delta = 1$ mm with no apparent cracking. Then, upon increasing $\delta$ to 1.5 mm, a crack initiated near the plastic zone boundary, as noted above. Further increases in $\delta$ extended the plastic zone ahead of the notch, and the crack propagated, approaching the center of the plastic zone. With increasing $\zeta$, cracks initiated at larger displacements: at $\delta \approx 2.5$ mm for $\zeta = 0.48$ (Figure 8B) and at $\delta \approx 4.5$ mm for $\zeta = 0.72$ (Figure 8C). In these composite cases, cracks typically initiated near the middle of the plastic zone. For the free-standing SEPS thin film, cracks did not initiate up to the highest displacement examined ($\delta = 4.5$ mm), although a crack did initiate at much higher displacements.

Similar behavior was observed in the center-hole-notched tensile specimens; however, as shown in Figure 6, crack initiation occurred only in the HDPE film. The displacement ($\delta$) at which cracks initiated is marked with red asterisks in the corresponding force plot, and the delay in crack initiation with an increasing rubber fraction ($\zeta$) is readily apparent.

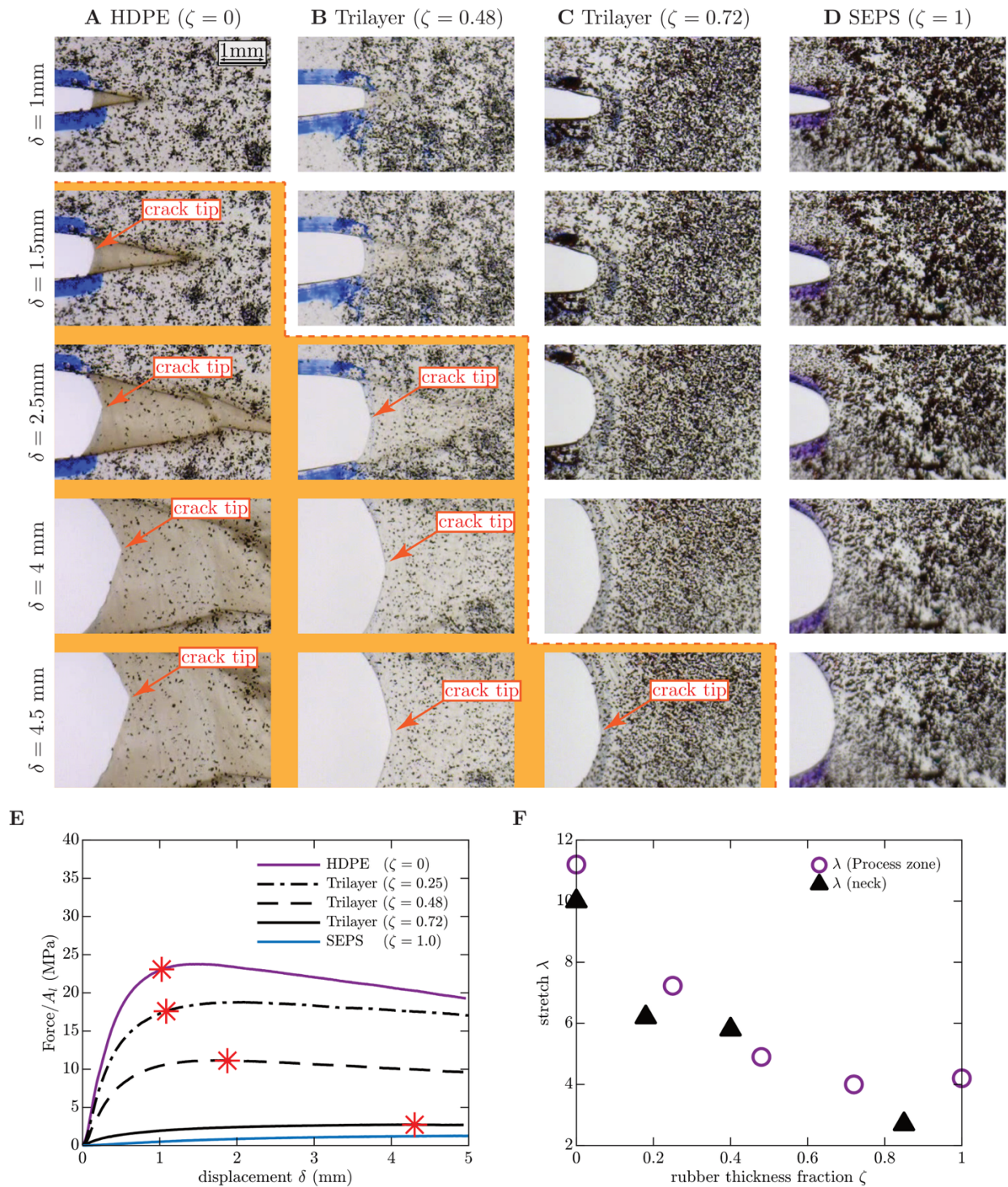

Figure 8: Plastic zone evolution in (A) HDPE, (B&C) HDPE-SEPS trilayer of $\zeta$ values 0.48 and 0.72 (D) SEPS. The displacement corresponding to each row is marked. Frames that show crack initiation and propagation are marked with orange background. (E) The normalized force-displacement plot. The crack initiation in each sample is marked using red asterisks. (F) Stretch ahead of crack tip is plotted with $\zeta$, measured at a $\delta = 4.5$ mm. The uniaxial neck stretches from Figure 4B is also plotted.

The high magnification permitted a quantification of the stretch inside the plastic zone. Stretch maps were calculated from video analysis. Figure 8F shows the measured stretch 0.5 mm away from the crack tip (within the plastic zone) when the applied displacement was $\delta = 4.5$ mm. The stretch ahead of the crack tip decreased with increasing $\zeta$. These values are in good agreement with the stretch of the necked region shown in Figure 4B.

### 4.5 Simulation of tensile deformation of defect-free HDPE and HDPE-SEPS-HDPE trilayers thin films

The following sections present the simulation results for defect-free and notched geometries under tension for HDPE and HDPE-SEPS-HDPE trilayers with various SEPS fractions ($\zeta$). We show the deformed configurations with increasing applied stretch ($\lambda_{app}$) using color maps corresponding to two quantities: the primary principal stretch ($\lambda_1$), as defined in Eq. 7, and the ductile damage parameter ($D$), as defined in Eq. 6. This section begins with the tensile simulation of defect-free rectangular specimens as described in Section 3.3. These simulations were performed to mimic the deformation in the gauge section of a tensile specimen, similar to the experiments shown in Figure 1C and 1D.

Figure 9A shows the simulated deformed configurations when stretching defect-free uniaxial tensile specimens with $\zeta = 0$ (HDPE), 0.25, 0.5, and 0.75 at specific applied stretch, $\lambda_{app} = 1.5, 1.9, 2.0$, and 2.25. The $\lambda_1$ colormap visualizes local stretches, and the $\lambda_{app}$ values were chosen to highlight neck propagation. For HDPE ($\zeta = 0$), the uniaxial simulation reproduces the necking and drawing behavior observed in experiments (Figure 1C). At $\lambda_{app} = 1.5$, the specimen stretches uniformly, creating a homogeneous $\lambda_1$ colormap. By $\lambda_{app} = 1.9$, a distinct neck has formed, and the colormap becomes heterogeneous. The stretch becomes further concentrated at $\lambda_{app} = 2.0$, with warmer colors signifying higher stretch concentrated within the neck. Finally, at $\lambda_{app} = 2.25$, the stretch in the neck stabilizes, and the neck boundaries begin to propagate along the specimen's length. Consistent with the experimental results (Figure 1C), the simulation captures a very sharp transition between the necked and un-necked regions, as highlighted by the abrupt color change in the magnified view.

Figure 9A also shows the deformed configuration and the $\lambda_1$ colormap when a SEPS layer is bonded to HDPE. Three major changes in behavior are apparent with the addition of rubber. First, increasing $\zeta$ delays the onset of necking (e.g., compare $\zeta = 0$ vs. 0.25 and 0.5). For a sufficiently large $\zeta$, necking is suppressed altogether (e.g., $\zeta = 0.75$). Second, the $\lambda_1$ in the necked region decreases with increasing $\zeta$, a trend visualized by the progressively cooler colors in the $\lambda_1$ colormap at the neck. For the fully suppressed case ($\zeta = 0.75$), the specimen stretches uniformly with $\lambda_1 = \lambda_{app}$ everywhere. Third, the transition between the necked and un-necked regions, which was sharp for HDPE, becomes more diffuse with increasing $\zeta$, as clearly illustrated by the smoother color transition in the magnified views.

Figure 9B, on the other hand, shows the same deformed configurations as in Figure 9A, but with the damage parameter $D$ plotted. Like the $\lambda_1$ field, the $D$ field is uniform until the onset of necking, and its highest value is at the center when necking initiates ($\lambda_{app} = 1.9$). However, at larger $\lambda_{app}$ values, unlike $\lambda_1$, which concentrated in the middle of the neck, the peak $D$ value moves to the edge of the necking zone. The zoomed-in view at $\lambda_{app} = 2.25$ shows the concentration of $D$ at the boundary of the neck. Furthermore, similar to $\lambda_1$, the maximum $D$ value decreases as $\zeta$ increases.

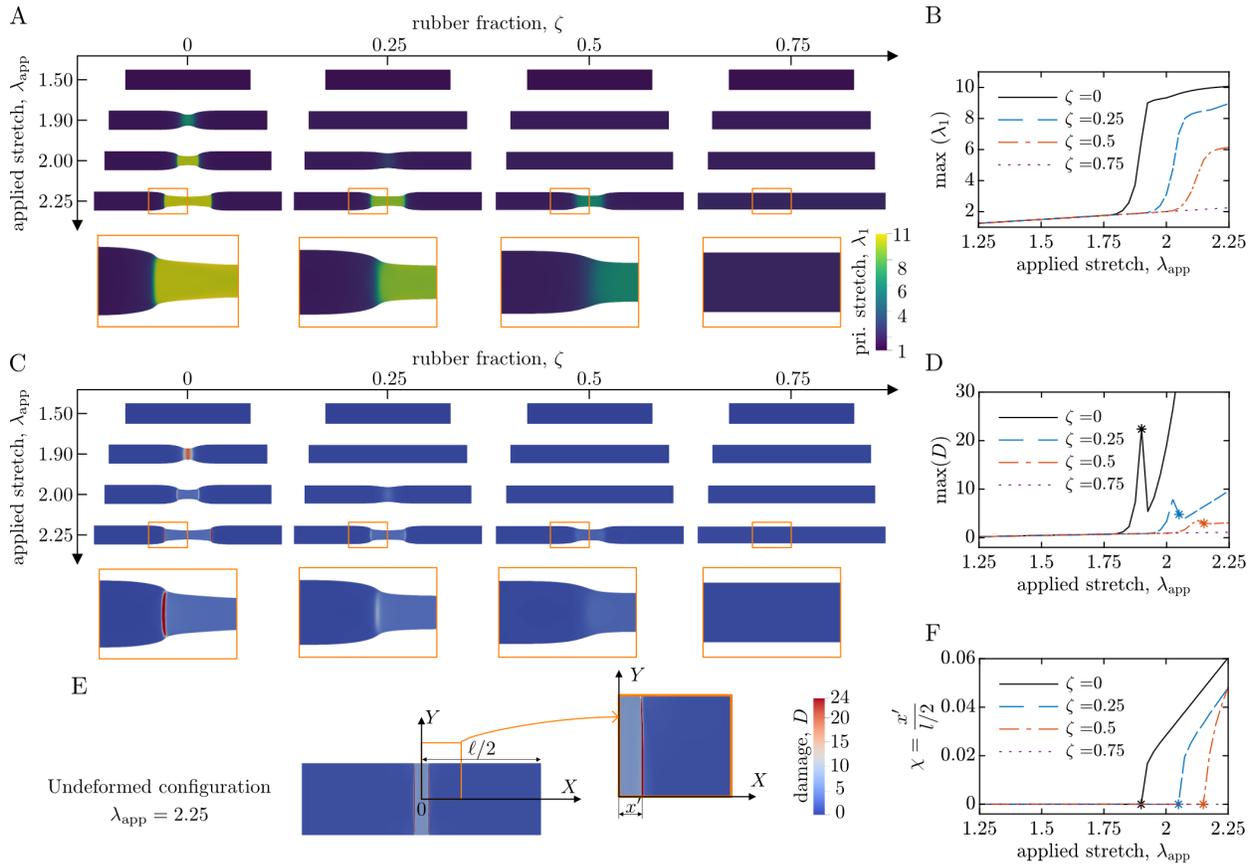

Figure 9: Primary principal stretch ($\lambda_1$) and ductile damage parameter ($D$) in a uniaxial tensile specimen. (A) Deformed configurations with the $\lambda_1$ field as the colormap for specimens with $\zeta = 0$ (freestanding HDPE), 0.25, 0.5, and 0.75, shown at applied stretch, $\lambda_{app} = 1.5$, 1.9, 2.0, and 2.25. Magnified views at $\lambda_{app} = 2.25$ highlight the region near the neck boundary. (B) The same deformed configurations as in (A) are plotted with the ductile damage parameter $D$ as the colormap (Eq. 6). (C) The maximum primary principal stretch, max($\lambda_1$), for the four specimens plotted against $\lambda_{app}$. (D) The maximum damage, max($D$), from the colormaps in (B) plotted against $\lambda_{app}$. (E) The normalized position of max($D$), $\chi = \frac{x'}{l/2}$, plotted against $\lambda_{app}$. (F) An illustration of the measurement of $x'$ and $l/2$ on the undeformed configuration of the HDPE tensile specimen corresponding to the state at $\lambda_{app} = 2.25$. Here, $x'$ is the position in the half-specimen such that $x' \in [0, l/2]$.

The behavior of Figure 9A and B is quantified in Figure 9 C-E. First, Figure 9C plots the maximum primary principal stretch, $\max(\lambda_1)$, against the applied stretch ($\lambda_{app}$) for all four specimens. This plot effectively tracks the evolution of the neck stretch, as $\lambda_1$ is homogeneous during uniform deformation and its maximum is always located at the neck's center after localization. For the specimens that exhibit necking ($\zeta = 0, 0.25$, and $0.5$), the curves show three distinct regimes: an initial gradual increase during uniform deformation, a sharp rise corresponding to strain localization and neck formation, and a final plateau as the neck stretch saturates and propagates. As $\zeta$ increases, the onset of necking is delayed (the jump occurs at a larger $\lambda_{app}$), and the plateau stretch value decreases. In contrast, for the $\zeta = 0.75$ specimen, in which necking is suppressed, $\max(\lambda_1)$ increases linearly with $\lambda_{app}$, confirming purely uniform deformation.

Next, Figure 9D and E quantify the damage evolution. Figure 9D plots the maximum damage, $\max(D)$, against $\lambda_{app}$, while Figure 9E tracks the normalized position of this maximum along the X-axis $\left(\chi = \frac{x'}{l/2}\right)$. The measurements of $x'$ and $l/2$ are taken from the undeformed configuration, as shown in Figure 9F. For the HDPE specimen ($\zeta = 0$, solid black curve), $\max(D)$ initially increases steadily while its position remains at the specimen's center ($\chi = 0$), signifying damage concentration in the forming neck. However, $\max(D)$ does not increase monotonically; it reaches a peak and then decreases. Concurrently, at a critical $\lambda_{app}$ (marked by an asterisk), the position $\chi$ becomes non-zero, indicating that as the $D$ value at the neck center decreases, the point of maximum damage moves from the neck's center to its boundary. This behavior is also visualized in the $D$ field plot in Figure 9B. With further stretching, $\max(D)$ begins to rise again, reflecting damage accumulation at the propagating neck front. For $\zeta = 0.25$ and $0.5$ (blue dashed and red dot-dashed curves), the qualitative behavior is similar, but the magnitude of $D$ is reduced. In contrast, for $\zeta = 0.75$ (dotted purple curve), the deformation remains homogeneous, so the damage $D$ is uniform and its magnitude remains small.

## 4.6 Simulation of tensile deformation of notched HDPE and HDPE-SEPS-HDPE trilayers thin films

We now analyze the $\lambda_1$ and $D$ distributions in notched specimens. We examine two types of notched geometries: a center-hole-notched tensile (CHNT) specimen and a pure shear (PS) specimen with a sharp edge notch. The results are presented in Figure 10 (CHNT) and Figure 11 (PS). For each geometry, we present the $\lambda_1$ and $D$ fields on the deformed configurations, plot the evolution of their maximum values ($\max(\lambda_1)$ and $\max(D)$) with $\lambda_{app}$, and track the location of the peak damage. The location of $\max(D)$ is quantified using the angular position $\theta$ from the notch tip, as illustrated in Figure 10F (CHNT) and Figure 11F (PS).

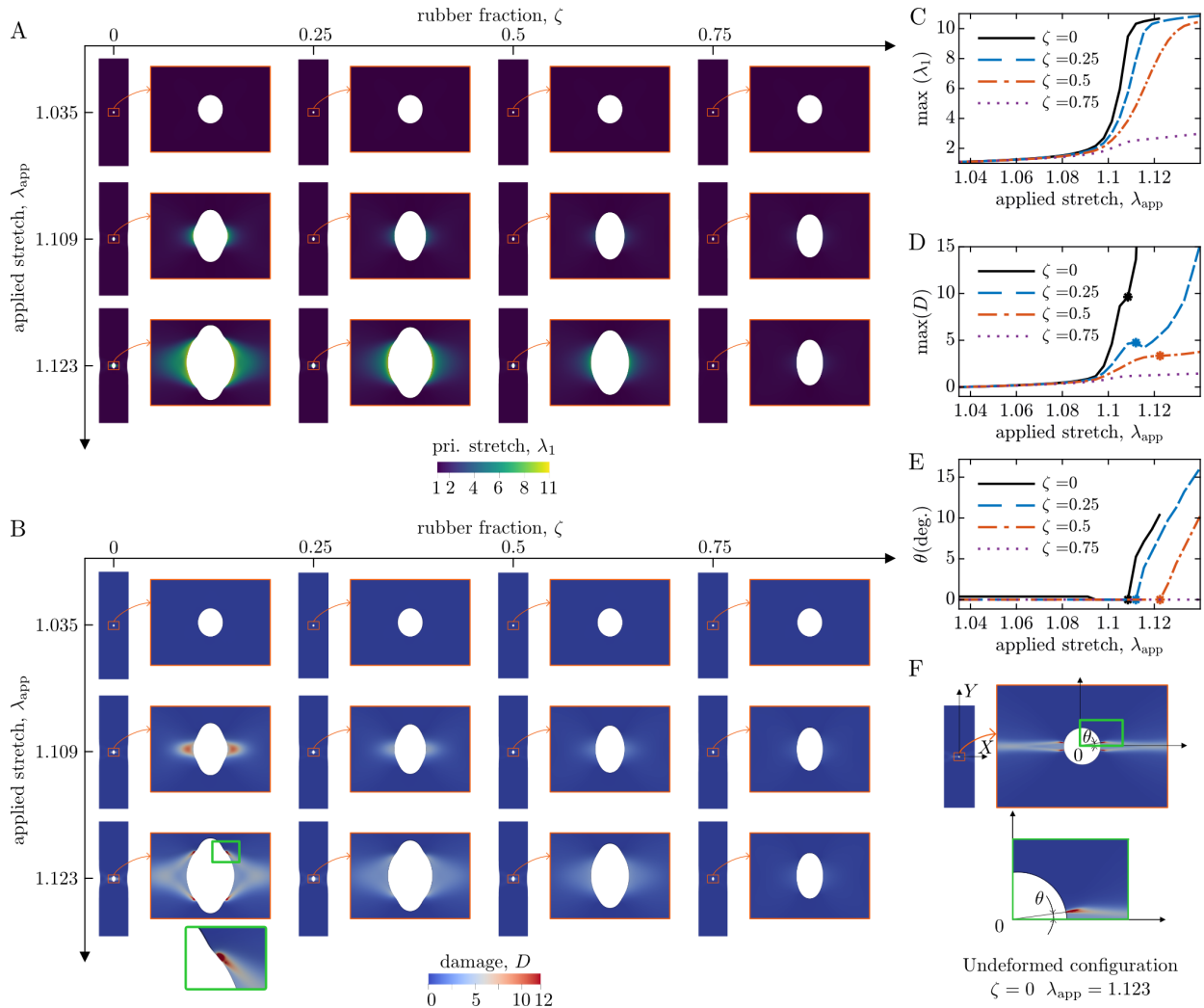

Figure 10: Primary principal stretch ($\lambda_1$) and ductile damage parameter ($D$) in center hole-notched tensile specimen. (A) Deformed configuration with $\lambda_1$ field as colormap for specimens with $\zeta$ values of 0 (Free-standing HDPE), 0.25, 0.5, and 0.75 respectively shown at applied stretch ($\lambda_{app}$) values of 1.035, 1.109, and 1.123. (B) The same deformed configuration in Figure 10A are plotted with colormap of $D$ (Eq. 6). (C) The max($\lambda_1$) for the four specimens in Figure 10A are plotted against $\lambda_{app}$. (D) The max($D$) from the colormaps in Figure 10B are plotted against $\lambda_{app}$. (E) The angular position of the max($D$) in the undeformed configuration $\theta$ (deg.) is plotted against $\lambda_{app}$. (F) An example of measurement of $\theta$, in the undeformed configuration of tensile center-hole-notched tensile specimen of HDPE at $\lambda_{app} = 1.123$. Here $\theta$ is the angular position in the quarter specimen such that, $\theta \in [0,90]$.

First, the primary principal stretch ($\lambda_1$) is shown in Figure 10A and Figure 11A, with its peak value quantified in Figure 10C and Figure 11C. For specimens that neck ($\zeta = 0$ (HDPE), 0.25, and 0.5), the behavior is similar to the defect-free tensile case, with a neck-like, high-stretch zone forming ahead of the notch. However, a key difference is that the notch localizes the stretch at its tip even for the non-necking trilayer ($\zeta = 0.75$). As with the tensile specimens, max($\lambda_1$) is always located at the horizontal mid-plane ($\theta = 0$), and its magnitude decreases with increasing $\zeta$.

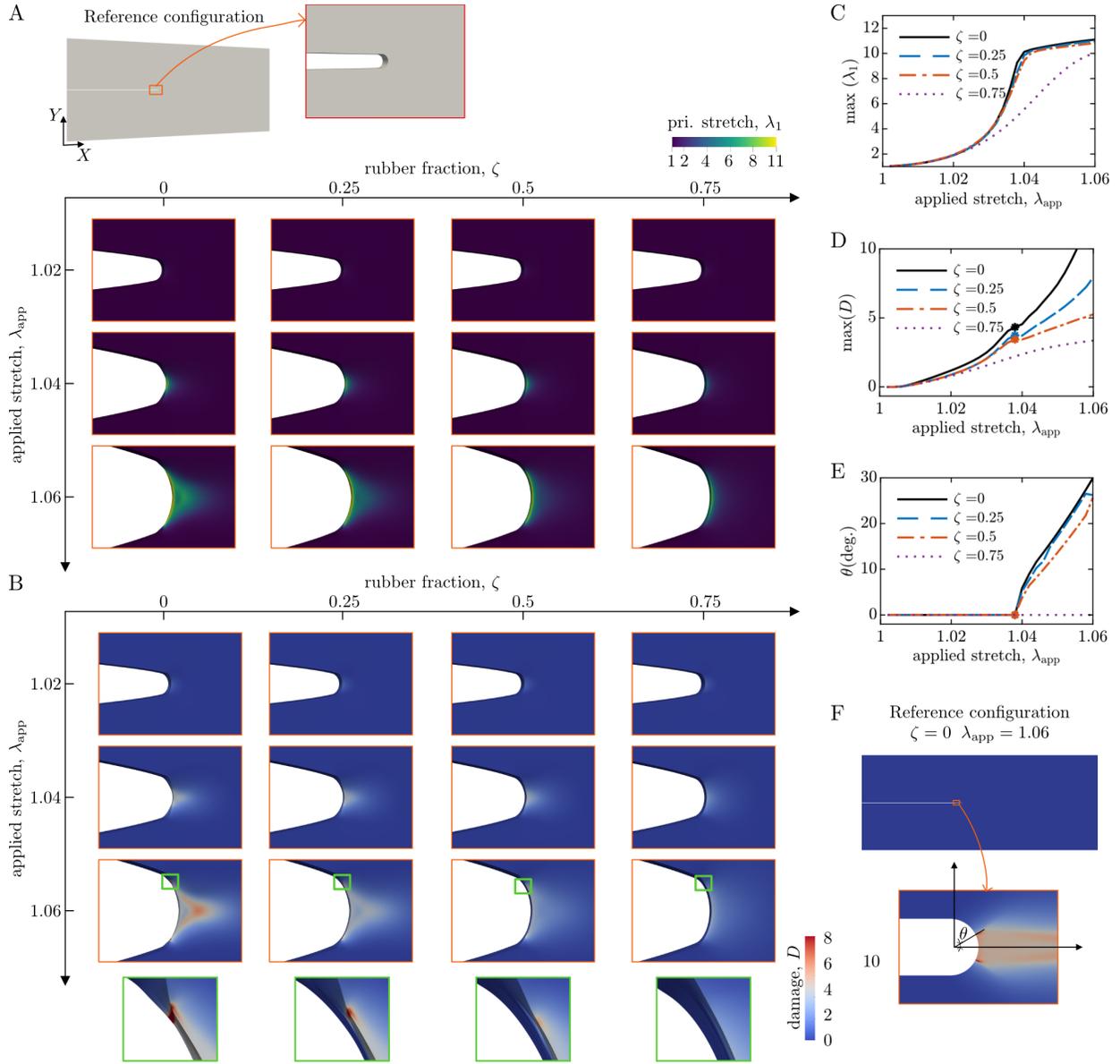

Figure 11: Primary principal stretch ($\lambda_1$) and ductile damage parameter ($D$) in center hole-notched tensile specimen. (A) Deformed configuration with $\lambda_1$ field as colormap for specimens with $\zeta$ values of 0 (Free-standing HDPE), 0.25, 0.5, and 0.75 respectively shown at applied stretch ($\lambda_{app}$) values of 1.02, 1.04, and 1.06. The specimen is tilted about the $Y$-axis and only a region of interest around the notch tip is shown in the illustration with undeformed specimen. (B) The same deformed configuration in Figure 11A are plotted with colormap of $D$ (Eq. 6). (C) The max($\lambda_1$) for the four specimens in Figure 11A are plotted against $\lambda_{app}$. (D) The max($D$) from the colormaps in Figure 11B are plotted against $\lambda_{app}$. (E) The angular position of the max($D$) in the undeformed configuration $\theta$ (deg.) is plotted against $\lambda_{app}$. (F) An example of measurement of $\theta$, in the undeformed configuration of pure shear fracture specimen of HDPE at $\lambda_{app} = 1.06$. Here $\theta$ is the angular position in the quarter specimen such that, $\theta \in [0,90]$.

Next, the ductile damage parameter ($D$) is shown in Figure 10B and Figure 11B. The evolution of $D$ is analogous to the defect-free tensile case. At small applied stretches, max($D$) is located at the notch tip ($\theta = 0$). As $\lambda_{\text{app}}$ increases, the location of the peak damage shifts from the notch tip to the edges of the high-stretch zone. This migration is quantified in the plots of the angular position $\theta$ versus $\lambda_{\text{app}}$ (Figure 10E and Figure 11E), where $\theta$ becomes non-zero at a critical applied stretch (marked by an asterisk). While the trend is similar, the drop in $D$ at the notch tip during this migration is less pronounced in the notched specimens than in the un-notched ones. As before, increasing $\zeta$ consistently reduces the overall magnitude of max($D$).

Comparing the notched geometries with the tensile specimen, we see that the qualitative behavior is similar, but the severity of the defect dictates the quantitative behavior. The sharp rise in $\lambda_1$ and $D$, as well as the migration of max($D$) away from the center (quantified by $\chi$ or $\theta$), occurs at progressively lower applied stretches as the stress concentration increases. Meanwhile, the reduction in max($D$) with $\zeta$ is largest for the tensile specimen, is lower for the center-hole-notched specimen, and is lowest for the pure shear fracture specimen. Furthermore, the effect of the $\zeta$ in reducing max($D$) differs at the two hotspots, being less pronounced at the neck center/notch tip than at the neck edge. For the PS fracture specimen, the reduction in max($D$) at the notch tip ($\theta = 0$) is minimal.

## 4.7 Effect of rubber strain hardening on damage parameter and maximum primary principal stretch

Our experiments used SEPS rubber as the middle layer of the composite; however, other rubber materials can also be used, with varying effectiveness. We now report on the effect of rubber strain hardening on the behavior of the trilayer for the three previously described specimens: tensile, center-hole-notched tensile, and pure shear fracture. For a trilayer with $\zeta = 0.25$, we increased the non-linear strain hardening parameter $C_3^r$ in the SEPS material model (Eq. 1) to 15 kPa, 50 kPa, and 500 kPa. Simulations were conducted on these modified trilayers with enhanced strain-hardening rubber, following the same procedures as those used for the results in Figure 9, Figure 10, and Figure 11. Our results, presented in Figure 12, demonstrate that increasing the rubber's strain hardening has a toughening effect similar to that of increasing the rubber fraction ($\zeta$). This enhanced hardening stabilizes the deformation and suppresses damage.

First, we analyze the maximum principal stretch, max($\lambda_1$), shown in panels A1, B1, and C1 of Figure 12. For the tensile specimen (Figure 12A1), increasing $C_3^r$ reduces the plateau stretch associated with necking, leading to more homogeneous deformation; at $C_3^r = 500$ kPa, necking is eliminated entirely. This homogenizing effect is less pronounced in the notched specimens (Figure 12B1 and Figure 12C1), where only the highest $C_3^r$ value produces a significant reduction in max($\lambda_1$).

Next, we analyze the damage evolution, which reveals a more complex trend, as the reduction in damage is dependent on its location. In the tensile specimen (Figure 12A2), increasing $C_3^r$ reduces the peak damage, max($D$), both at the neck's center ($\chi = 0$) and its boundary ($\chi > 0$). However, the reduction is far more pronounced at the boundary. This effect is amplified in the notched specimens (Figure 12B2 and Figure 12C2), where increasing $C_3^r$ has a minimal effect on the damage at the immediate notch tip ($\theta = 0$) but is more effective at suppressing damage in the plastic zone boundary ($\theta > 0$).

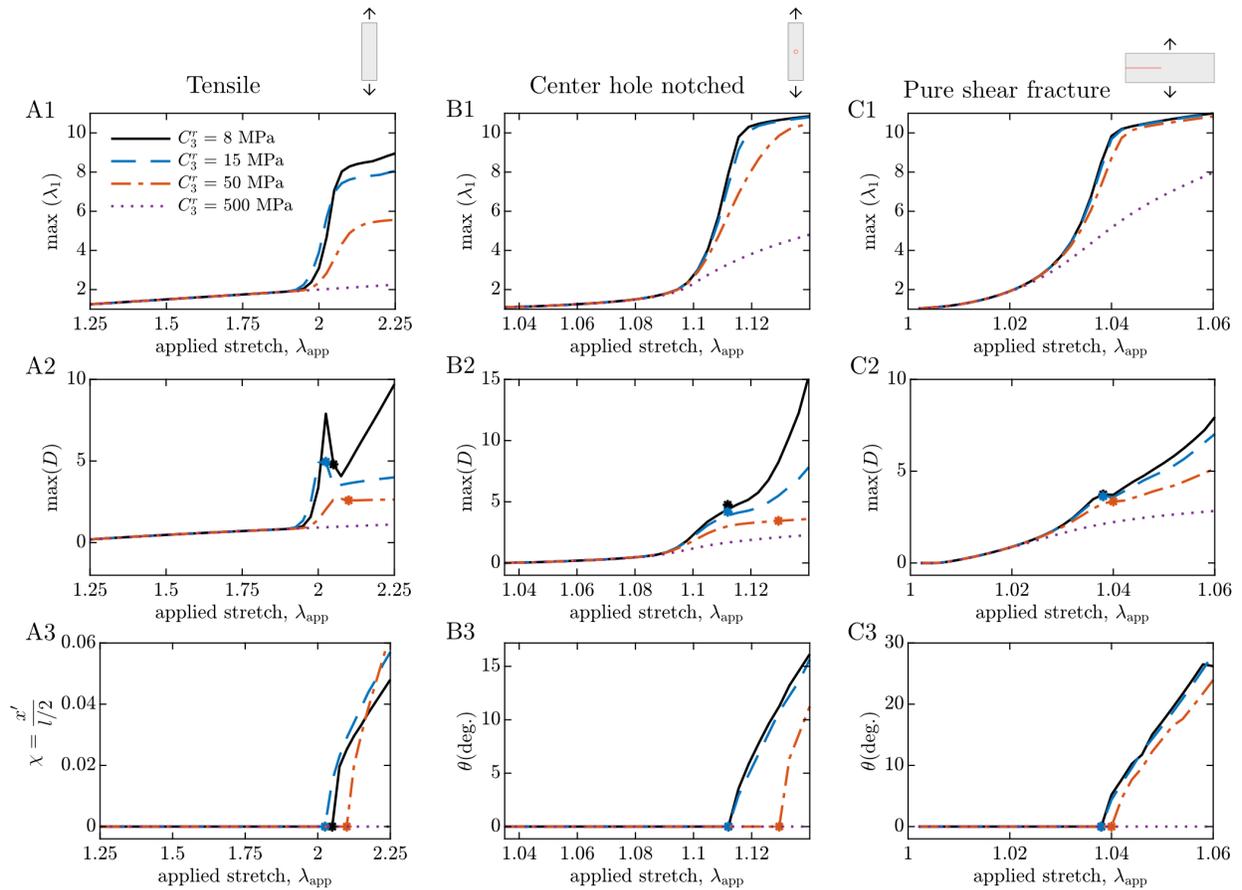

Figure 12: Effect of rubber strain hardening parameter $C_3^r$ on max($\lambda_1$) (top row), max($D$) (middle row) as well as: $\chi$ or $\theta$ (bottom row) in a trilayer of $\zeta = 0.25$ for (A) tensile specimen, (B) center-hole-notched specimen, and (C) pure shear specimen. The plots are made at $C_3^r$ values of 8 kPa (black solid curve), 15kPa (blue dashed curve), 50 kPa (red dot-dashed curve), and 500 kPa (purple dotted curve). The $C_3^r = 8$ kPa plots are reproduced from Figure 9, Figure 10, and Figure 11 respectively.

Finally, a comparison across the three geometries reveals a clear conclusion. The ability of rubber strain hardening to mitigate strain and suppress damage diminishes with increasing stress concentration. The toughening effect is most pronounced in the defect-free tensile specimen, less effective in the center-hole-notched specimen, and least effective in the pure shear specimen.

## 5 Discussion

This study was motivated by a key observation: a freestanding HDPE thin film fails prematurely via fracture at the neck boundary, whereas an HDPE-SEPS-HDPE trilayer composite undergoes stable, extensive stretching. This suggests that while HDPE is inherently ductile, it suffers from a flaw intolerance that can be mitigated by a soft elastomeric interlayer. The flaw tolerance map in Figure 5 for notched tensile specimens (6mm wide), confirmed that free-standing HDPE is extremely flaw intolerant for both the notch geometries, with a normalized flaw tolerance limit of $(c/w)^{HDPE} < 0.016$ for edge notches and $(d/w)^{HDPE} < 0.019$ for center hole notches. Meanwhile free-standing SEPS is sensitive to flaw size and shape, with a flaw tolerance limit of $c/w^{SEPS} \approx 0.05$ for the edge notches and $d/w^{SEPS} > 0.125$ for the center hole notches. Incorporating a flaw sensitive SEPS interlayer significantly enhanced the tolerance of HDPE. This improvement increases with the rubber's thickness fraction ($\zeta$), though the composite's maximum tolerance is ultimately bounded by that of the freestanding SEPS layer itself.

This discussion will first elucidate the specific ductile fracture mechanism governing failure in free-standing HDPE and then explain how the addition of a rubber layer fundamentally alters this mechanism to enhance toughness and flaw tolerance.

### 5.1 HDPE thin film undergoes ductile fracture

Before we discuss the specific crack initiation and failure mechanism in HDPE thin films, it is useful to highlight the failure mode in HDPE films. Material failure occurs when local stresses exceed local microscopic structural limits, and failure becomes energetically favorable (Anderson, 2017a). The stress state also affects the failure strain. Stress triaxiality (Eq. 5) reduces crack initiation strain in both metals (Bao, 2005; Gao et al., 2010; Hancock and Mackenzie, 1976; Johnson and Cook, 1985; Mackenzie et al., 1977; McClintock, 1968; Needleman and Tvergaard, 1984) and polymers (Wang et al., 2000). In contrast, yielding causes crack tip blunting and relives stress triaxiality and consequently is generally associated with ductility in materials (Anderson, 2017b). Furthermore, in ductile materials, yielding influences fracture behavior by dissipating energy through plastic deformation (Anderson, 2017b). Hence the interplay between yielding and stress triaxiality governs the brittle-to-ductile transition in materials. In this context, the brittle-to-ductile transition in HDPE has received considerable attention (Kitao, 2001; Krishnaswamy, 2005; O'connell et al., 2002; Ognedal et al., 2014).

While yielding in HDPE is sensitive to temperature and loading rate, stress triaxiality is sensitive to geometry. Yield stress decreases with temperature (Lamri et al., 2020; Ward and Sweeney, 2012) and increases with loading rate (Dasari and Misra, 2003; Ward and Sweeney, 2012). Meanwhile, brittle-to-ductile transition occurs with decreasing specimen thickness and increasing notch radius in polycarbonate, another stiff-soft-stiff polymer that exhibits neck propagation (Kattekola et al., 2013). Although the failure of HDPE thin film tensile specimens (as shown in Figure 1C) and all notched specimens discussed here (Figure 6 A, Figure 7 A, and Figure 8A) occurs at relatively small applied stretches, the formation of necked zones and significant plastic deformation confirms that the failure mode is not brittle fracture even for the geometry with sharp notches. Additionally, the load curves during failure do not show an abrupt drop, usually associated with brittle failure, as shown in Figure 4C for tensile specimen and Figure 7 B for fracture specimen. Taken together, we see that the HDPE thin film is a fully ductile material in Orowan's brittle-ductile classification(Orowan, 1949; Ward and Sweeney, 2012). The three possible classifications are: first, a brittle material; second, a notch-brittle material which is ductile when specimen is unnotched and brittle when a sharp notch is introduced; and third, a fully ductile material, where the material is ductile for unnotched as well as notched specimens. Therefore, a ductile fracture criterion used is for HDPE thin films across all three geometries studied, i.e., the tensile specimen, the CHNT specimen and the PS fracture specimen.

**5.2 HDPE fractures at the second hotspot for ductile damage parameter *D***

Our experiments on free-standing HDPE films consistently show that in all tested geometries - uniaxial tension, CHNT, and PS fracture specimens - the crack initiates not at the point of maximum stretch (i.e., the center of the neck or notch tip) but at the boundary of the neck-like plastic zone (see tensile specimen in Figure 1C, CHNT specimen in Figure 6A, and PS fracture specimen in Figure 8A). It is useful to compare this observation with another material system. Polycarbonate is a much-studied stiff-soft-stiff, notch-brittle material. Just like HDPE, the polycarbonate also forms highly stretched, neck-like plastic zone at the notch tip. Under high stress-triaxiality conditions, the brittle failure is found to initiate ahead of the plastic zone tip (Ishikawa et al., 1977; Narisawa and Yee, 2006); hence, a hydrostatic stress-based criterion has been used to characterize failure (Ishikawa et al., 1977; Nimmer and Woods, 1992). Meanwhile under low stress triaxiality condition, ductile fracture initiates at the notch tip, inside the plastic zone. Here, a plastic strain-based criterion has been found suitable to characterize failure (Gearing and Anand, 2004).

Since the HDPE thin film is fully ductile, we do not consider a purely hydrostatic stress-based condition for HDPE. Furthermore, finite element simulation results (Figure 9, Figure 10, and Figure 11) reveal that cracks do not initiate at the location of maximum principal stretch or plastic strain, i.e., the middle of the neck and the notch tip. Therefore, we rule out maximum principal stretch as a suitable failure criterion for HDPE thin films.

The ductile damage parameter $D$, presented in Eq. 6, is adapted from the Hancock and Mackenzie failure criterion (Hancock and Mackenzie, 1976; Wang and Kuang, 1995), using the void growth rate derived by Rice and Tracey (Rice and Tracey, 1969). The parameter $D$ scales exponentially with stress triaxiality $\eta$ and linearly with plastic strain $\epsilon_p$. Higher values of $D$ indicate an increased rate of void growth, and thus a greater propensity for crack initiation. Similar approach has been used to characterize the ductile fracture in metals (Garrison and Moody, 1987; Rousselier, 1987) such as steel (Sowerby and Chandrasekaran, 1986) and aluminum alloys (Bao, 2005; Bao and Wierzbicki, 2004) as well as polymers like polycarbonate (Wang et al., 2000) and semicrystalline polymer polypropylene (Wiersma and Sain, 2023).

Our finite element simulations, using this ductile damage parameter ($D$), revealed a critical insight: for all geometries, two "damage hotspots" emerge. The first hotspot forms at the location of maximum stretch (the notch tip or neck center) early in the deformation. However, as deformation proceeds, a second, more intense hotspot develops at the boundary of the plastic zone (see Figure 9B, D, Figure 10B, D, and Figure 11B, D). It is this second hotspot that corresponds to the experimentally observed crack initiation sites. This strongly suggests that failure in HDPE thin films is governed by the accumulation of ductile damage, which is maximized in the notch face at the interface between the highly deformed neck and the surrounding bulk material, due to the combination of the localized peak in stress triaxiality and plastic strain. Our conclusion is consistent with previous experimental work showing that void growth in HDPE increases with stress triaxiality (Ben Jar and Muhammad, 2012; Mesbah et al., 2021; Ognedal et al., 2014), and is further supported by visual evidence of discoloration - indicative of intense voiding - in this exact region (Figure 13).

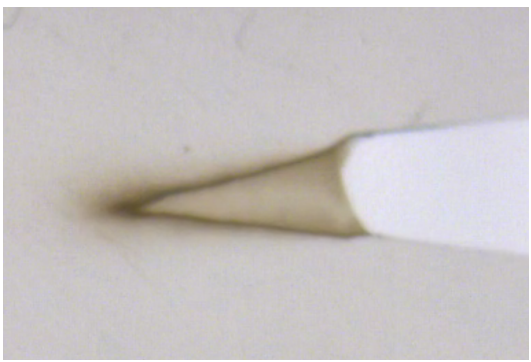

Figure 13: Discoloration, possibly resulting from void nucleation within the highly stretched neck-like plastic zone and its boundary, observed at the notch tip of the HDPE pure shear fracture specimen.

### 5.3 Rubber layer alters failure mechanism in HDPE-SEPS-HDPE laminates

The addition of a soft SEPS interlayer dramatically improves the material's performance by fundamentally altering the failure mechanism. Crucially, the crack initiation is both delayed and relocated to the notch tip. This effect is evident in the center-hole-notched tensile experiment (Figure 6 B) as well as in the pure shear fracture experiment (Figure 8 B). Meanwhile, from the simulations we see that addition of rubber reduces the magnitude of peak damage, $\max(D)$ and maximum primary principal stretch, $\max(\lambda_1)$ in tensile specimen (Figure 9 C and D, respectively), CHNT specimen (Figure 10 C and D, respectively), and PS fracture specimen (Figure 11 C and D, respectively).

We also note that addition of SEPS layer has a larger effect in reduction of $\max(D)$, as opposed to reduction of $\max(\lambda_1)$. This effect is even more dominant in the CHNT specimens (Figure 10) and PS fracture specimens (Figure 11) where very little reduction of $\max(\lambda_1)$ is observed concurrent to the significant reduction in $\max(D)$.

With the $D$ driven failure mechanism suppressed, failure becomes governed by a different mode: a critical stretch ($\lambda_1$) criterion. The maximum stretch remains concentrated at the notch tip, which now becomes the weakest point and the site of crack initiation. The failure may originate in the SEPS layer, as the stretch at the notch tip is significantly higher than that in a free-standing SEPS layer at the same applied stretch, $\lambda_{\text{app}}$ (see Figure 8 F as well as Figure 9 C, Figure 10 C, and Figure 11 C). Furthermore, the $\max(\lambda_1)$ decreases with increasing $\zeta$ for all the three geometries, which explain the experimentally-observed increase in $\lambda_{\text{app}}$ for crack initiation with increasing $\zeta$ in HDPE-SEPS-HDPE laminates (see Figure 8).

### 5.4 Tuning failure in HDPE-SEPS-HDPE laminates with rubber layer properties

We have shown that the failure in HDPE can be tuned by controlling two parameters, $\max(D)$ and $\max(\lambda_1)$. We have presented two pathways to achieve this; first varying the thickness of the rubber layer (see Figure 9 for tensile specimens, Figure 10 for center-hole-notched tensile specimens, and Figure 11 for fracture specimens) and second by varying the rubber strain hardening (Figure 12). Increasing the rubber thickness fraction $\zeta$ and rubber strain hardening $C_3^r$ have qualitatively similar effect on $\max(D)$ and $\max(\lambda_1)$.

The effect of rubber layer on these two parameters depends on the specimen geometry. Specifically, the ability of $\zeta$ and $C_3^r$ to reduce $\max(D)$ is most pronounced in the tensile specimen (Figure 12 A2 and Figure 9D), but its effectiveness decreases in the CHNT specimen (Figure 12 B2 and Figure 10D), and is even less pronounced in the PS fracture specimen (Figure 12 C2 and Figure 11D). This effect explains the discrepancy in improving stretchability with the addition of a SEPS layer experimentally observed in tensile specimen (Figure 4C), CHNT specimen (Figure 5B), and ENT specimens (Figure 5A). Bonding a thin SEPS layer was sufficient to allow stretching without failure in tensile specimen (Figure 1D), while CHNT showed significant jump in stretchability. In contrast, ENT specimens became flaw-sensitive but gained little by increasing SEPS fraction $\zeta$ Figure 5A).

# 6 Summary and relevance

In summary, this work identifies a critical, previously overlooked failure mode in HDPE thin films driven by ductile damage accumulation at the neck boundary. Experiments showed that HDPE thin films fail by the formation of a neck in the tensile specimen, followed by crack initiation at the boundary of the necked region. In notched specimens, the failure process involves the formation of a highly stretched, neck-like plastic zone that blunts the notch, which is then followed by crack initiation near the boundary of the plastic zone. The addition of SEPS prevented the failure of HDPE during tensile deformation in defect-free specimens. In notched specimens, the crack initiation strain increased to large applied displacements with an increasing SEPS fraction, for both CHNT specimens and PS fracture specimens. Concurrently, the crack initiation location changed from near the plastic zone boundary in HDPE to the middle of the plastic zone in the laminates.

Finite element simulations suggest that void-growth-driven ductile damage ($D$) is the driving mechanism for failure in HDPE. The addition of rubber offers toughening by lowering the magnitude of ductile damage. The effectiveness of rubber in damage reduction also depends on the geometry, with decreasing effectiveness for the center-hole notch and, subsequently, the sharp edge notch. Furthermore, the change in failure location from the plastic zone boundary to the notch tip is due to the change in failure mechanism from a ductile-damage-based mechanism in the HDPE to a stretch-based mechanism in the rubber. This finding provides a mechanistically-informed pathway for designing tougher, more reliable HDPE composites by tuning the properties and geometry of the rubber layer to control both damage and stretch localization.

The principles identified here may have broader relevance. The ductile fracture in polycarbonate initiates at the notch tip (Gearing and Anand, 2004), this coincide with the location of initial damage concentration in HDPE. Furthermore, just as elastomer interlayer improved stretchability of HDPE, previously elastomeric substrate has found to improve stretchability in metals (Li et al., 2004). Apart from the toughening effects, this work highlights the role of ductile damage in post yield fracture. Recent in-situ observation of voids in neck of a mild steel uniaxial specimen shows that the largest void volume coincide with the region of largest stress triaxiality, which is not at the center of the neck but towards the transition region of the neck (Zhang et al., 2023). These connections suggest that the interplay between ductile damage and stretch localization in the plastic zone may be a key, and often overlooked, factor in the fracture of many ductile material systems.

## Acknowledgements

This research was supported by the grants NSF-CMMI 1636064, NSF-2036164 and NSF-CMMI-1561789.

This material is based upon work supported by (while serving at) the National Science Foundation. Any opinion, findings, and conclusions or recommendations expressed in this material are those of the author(s) and do not necessarily reflect the views of the National Science Foundation or of the Federal government.

During the preparation of this work the authors used generative AI tool Google Gemini in order to improve Grammar and readability in the abstract, introduction, results, and discussion. After using this tool, the authors reviewed and edited the content as needed and take full responsibility for the content of the publication.

# Supplementary Information for Ductile fracture of HDPE thin films: failure mechanisms and tuning of fracture properties by bonding a rubber layer

## Deformation at notch tip of an edge notched tensile specimen

HDPE specimen formed a triangular plastic zone; later stage deformation was accommodated entirely by drawing new material into the plastic zone; eventually a crack initiated near the boundary of the plastic zone. The SEPS specimen showed no distinctly-localized plastic zone and no cracks developed. Composite showed intermediate behavior: the plastic zone had less distinct boundaries; deformation was accommodated by both recruiting new material or by stretching existing material in the plastic zone (depending on rubber content); the cracks started near the boundary of the plastic zone at low rubber content and near the center for high rubber content.

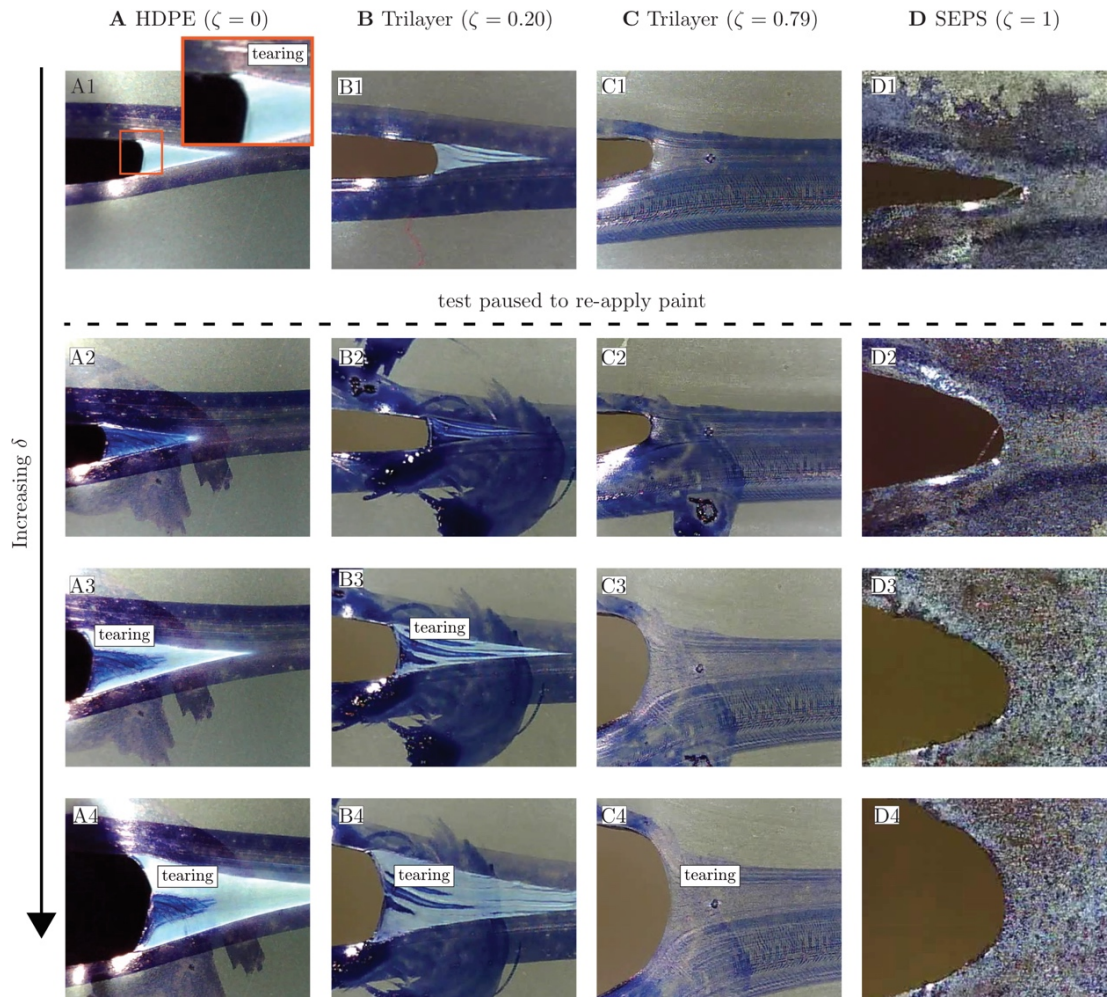

Figure S1: Plastic zone evolution in (A) HDPE, (B&C) HDPE-SEPS trilayer of $\zeta$ values 0.20 and 0.79, and (D) SEPS rubber, captured using a two-step experiment. First row shows the formation of plastic zone ahead of the crack. The test is then paused, and paint is reapplied on the material in the plastic zone (row 2). Evolution in the plastic zone with further stretching is shown in row 3 and row 4.